\documentclass[prl,twocolumn,floatfix,showpacs,preprintnumbers,superscriptaddress,amsmath,amssymb]{revtex4}
\usepackage[dvips]{graphics}
\usepackage{psfrag}
\usepackage{graphicx}
\usepackage{ucs}
\usepackage[utf8]{inputenc}

\begin{document}

\title{Co-tunneling current and shot noise in quantum dots} 

\author {Axel Thielmann} 
\affiliation{Forschungszentrum Karlsruhe, Institut f\"ur Nanotechnologie,
76021 Karlsruhe, Germany}

\author {Matthias H. Hettler} 
\affiliation{Forschungszentrum Karlsruhe, Institut f\"ur Nanotechnologie,
76021 Karlsruhe, Germany}

\author {J\"urgen K\"onig}
\affiliation{Institut f\"ur Theoretische Physik III, Ruhr-Universit\"at
             Bochum,  44780 Bochum, Germany}

\author {Gerd Sch\"on} 
\affiliation{Forschungszentrum Karlsruhe, Institut f\"ur Nanotechnologie,
76021 Karlsruhe, Germany}
\affiliation{Institut f\"ur Theoretische Festk\"orperphysik,
Universit\"at Karlsruhe, 76128 Karlsruhe, Germany}

\date{\today}

\begin{abstract}
We derive general expressions for the current and shot noise, 
taking into account non-Markovian memory effects. In generalization of previous
approaches our theory is valid for arbitrary Coulomb interaction and 
coupling strength and is applicable to quantum dots and more complex 
systems like molecules.
A diagrammatic expansion up to second-order 
in the coupling strength, taking into account co-tunneling
processes, allows for a study of transport in a regime relevant to many 
experiments.
As an example, we consider a single-level quantum dot, focusing 
on the Coulomb-blockade regime. We find 
super-Poissonian shot noise due to spin-flip 
co-tunneling processes at an energy scale different from the one expected 
from first-order calculations, with a sensitive dependence on  the 
coupling strength. 

\end{abstract}

\maketitle

{\bf Introduction.} --
The study of shot noise in transport through mesoscopic devices, such
as quantum dots or molecules,  
has become a field of intense theoretical and experimental research\cite{blanter}.
It provides  additional 
information, not contained in the current,
about system parameters that govern the 
electronic transport\cite{thielmann}.
For weak coupling between the  dot
and metallic electrodes, transport is 
dominated by sequential tunneling, usually described by first-order 
perturbation theory in the coupling strength.
Shot noise in this limit has been studied for a variety of models, including
effects from electronic interaction\cite{thielmann,sequential},  multi-level 
dots\cite{bulka,thielmann_cm}, inclusion of
photonic and vibrational modes\cite{thielmann_cm,aleiner}, or spin-polarized 
leads\cite{ferro_leads}. However, a 
theory taking fully into account interaction effects as well as
arbitrary coupling strength has not been presented.
Expressions for the shot noise in terms of non-equilibrium 
Green functions have been derived only in the
absence of Coulomb interactions\cite{balatsky} or in a perturbative 
expansion thereof\cite{hershfield,hamasaki}.

Shot noise has been measured in various experimental 
realizations\cite{birk,savchenko,nauen}. As the strength of the coupling in a
given experiment is not {\it a priori} known, 
it is unclear whether first order calculations are sufficient.
Second-order tunneling
(co-tunneling\cite{averin}) processes can play an important
role for the conductance\cite{kouwenhoven}, particularly in the
Coulomb-blockade regime. Shot noise in this regime has been discussed 
 in Ref.~\onlinecite{loss} where the possibility of 
super-Poissonian noise was suggested. 
However, that work was limited to pure co-tunneling 
processes, which is too restrictive for the experiment of
Ref.~\onlinecite{kouwenhoven}.

In this Letter we provide a general formulation of the 
current and shot noise, valid for arbitrary coupling strength, 
while accounting fully for the  Coulomb interaction.
By making use of a diagrammatic formulation we expand 
the expressions 
up to second order in the coupling strength  and study
as an example a single-level quantum dot, which is 
out of equilibrium due to an applied bias voltage.
The theory covers the Coulomb-blockade regime (low bias), the sequential 
tunneling regime (bias larger than the first single-charge excitation
energy) and the crossover between both.
We find super-Poissonian shot noise in the Coulomb-blockade 
regime due to spin-flip co-tunneling processes 
at a different energy scale than expected from first-order calculations. 
Furthermore, we show that the noise to current ratio 
is highly sensitive to the coupling strength.
This may serve as an additional spectroscopic tool for the couplings.

{\bf The model.} --
The Anderson impurity model is based on the Hamiltonian
$\hat H = \hat H_{\rm L} + \hat H_{\rm R} + \hat H_{\rm dot} + 
\hat H_{\rm T,L} + \hat H_{\rm T,R}$ with 
$\hat H_{r} = \sum_{k \sigma}\epsilon_{k \sigma r} a_{k \sigma r}^{\dag}
  a_{k \sigma r}$, $\hat H_{\rm dot} = \sum_{\sigma}\epsilon_{\sigma}
  c_{\sigma}^{\dag}   c_{\sigma} + U n_{\uparrow}n_{\downarrow}$ and 
$\hat H_{{\rm T},r}= \sum_{k \sigma}(t_r a_{k \sigma r}^{\dag} c_{\sigma} 
+ h.c.)$ for $r={\rm L},{\rm R}$.
Here, $\hat H_{\rm L}$ and $\hat H_{\rm R}$ describe the left and right 
electrode with non-interacting electrons, $\hat H_{\rm dot}$ the 
quantum dot  with one (spin-dependent) level of energy 
$\epsilon_{\sigma}$ and Coulomb interaction $U$ for double occupancy
of the dot.
Tunneling between the leads and the dot is modeled by 
$\hat H_{\rm T,L}$ and $\hat H_{\rm T,R}$.
The coupling strength is characterized by the intrinsic line width 
$\Gamma_r = 2\pi \rho_e |t_r|^2$, and 
$\Gamma = \Gamma_{\rm L} + \Gamma_{\rm R}$, where
$\rho_e$ is the (constant) 
density of states of the leads.
The creation operators $a_{k \sigma r}^{\dag}$ and
$c_{\sigma}^{\dag}$ refer to the 
electrodes and the impurity (dot), and $n_\sigma = c_{\sigma}^{\dag} c_{\sigma}$.

We are interested in the current $I$ and the (zero-frequency) current noise
$S$, which for $eV \gg k_{\rm B} T$ is dominated by shot noise.
Both are  related to the (symmetrized) current operator 
$\hat I = (\hat I_{\rm R} - \hat I_{\rm L})/2$ with $\hat{I}_r = -i(e/\hbar) 
\sum_{k \sigma} \left( t_r a_{k \sigma r}^{\dag} c_{\sigma} -
h.c.\right)$, from which we obtain  $I = \langle \hat{I} \rangle$ and 
\begin{equation}
S = 2 \int_{-\infty}^0 dt \, \left[ \langle \hat I(t)\hat I(0)+
\hat I(0)\hat I(t) \rangle - 2 \langle \hat I \rangle^2 \right] \, .
\label{noisedef}
\end{equation}

{\bf Diagrammatic technique.} --
In Ref.~\onlinecite{thielmann} we formulated a theory of current noise 
for transport to first order in $\Gamma$, based on a diagrammatic language 
that has been developped for a systematic perturbation expansion of 
the current through localized levels\cite{diagrams}.
All transport properties are governed by the non-equilibrium time evolution of
the density matrix.
The electrode degrees of freedom can be integrated out, and
we obtain a reduced density matrix for the dot degrees of
freedom only, labeled by $\chi$. 
The time evolution of the reduced density matrix, described by the propagator
$\Pi_{\chi' \chi}(t',t)$ for the propagation from a state $\chi$ at time $t$ to
a state $\chi'$ at time $t'$, can be visualized by
diagrams\cite{diagrams,thielmann} on the Keldysh contour.
The full propagation is expressed as a sequence of irreducible blocks 
(self energies) $W_{\chi' \chi}(t',t)$ that are associated with transitions 
from state $\chi$ at time $t$ to state $\chi'$ at time $t'$.
This yields the Dyson equation 
\begin{equation}
  {\bf \Pi}(t',t) = {\bf 1} + \int_t^{t'} dt_2 \int_{t}^{t_2} dt_1 \,
  {\bf W}(t_2,t_1) {\bf \Pi}(t_1,t)
\label{eq:dyson}
\end{equation}
for the propagator, where the bold face indicates matrix notation related to 
the dot state labels.
For the following, it is convenient to introduce the Laplace transform
${\bf W}(z) = \hbar \int_{-\infty}^0 dt\, e^{z t}{\bf W} (0,t)$ and the
definitions ${\bf W} = {\bf W}(z)|_{z=0^+}$ and $\partial {\bf W} = 
\left( \partial {\bf W}(z)/\partial z \right)|_{z=0^+}$.

In the long-time limit, i.e. for time differences $t'-t$ larger than the 
correlation time ($\propto 1/\Gamma$) over which the system forgets its 
initial state, the propagator becomes ${\bf p}^{\rm st} \otimes {\bf e}^T$, 
where
${\bf e}^T = (1,\ldots ,1)$, and ${\bf p}^{\rm st}$ is the vector of the 
stationary probabilities, determined from
${\bf W} {\bf p}^{\rm st} = \bf 0$, independent of the initial 
(diagonal) density matrix.
As ${\bf W}$ has a zero eigenvalue, it cannot be inverted.
With the  normalization condition ${\bf e}^T {\bf p}^{\rm st} = 1$
we obtain the stationary probabilities ${\bf p}^{\rm st}$ 
by solving
\begin{equation}
  \left( {\bf \tilde W} {\bf p}^{\rm st} \right)_\chi 
  = \Gamma \delta_{\chi, \chi_0}  \, ,
\label{eq:master}
\end{equation}
where $\bf \tilde W$ is identical to $\bf W$ but with one (arbitrarily chosen) 
row $\chi_0$ being replaced by $(\Gamma,\ldots,\Gamma)$\cite{thielmann}.
For a diagrammatic representation of the current, we introduce a block 
${\bf W}^I$, in which one (internal) vertex is replaced by an
external due to the current operator. 
We get
\begin{equation}
  I = {e\over 2\hbar} {\bf e}^T {\bf W}^{I} {\bf p}^{\rm st} \, .
\label{eq:Ik}
\end{equation}

The shot noise, Eq.~(\ref{noisedef}), involves expectation values of two 
current operators.
They either appear both in a single irreducible 
block, which we denote by ${\bf W}^{II}$, or in two different blocks 
${\bf W}^{I}$.
We find
\begin{equation}
  S = {e^2\over \hbar} {\bf e}^T \left[ 
    {\bf W}^{II} + {\bf W}^{I} ( {\bf P} {\bf W}^{I} 
    +{\bf p}^{\rm st} \otimes {\bf e}^T \partial {\bf W}^{I} ) \right]
    {\bf p}^{\rm st} \, 
\label{eq:noise_ptilde}
\end{equation}
with $\partial {\bf W}^{I} = 
\left( \partial {\bf W}^{I}(z)/\partial z \right)|_{z=0^+}$.
The object ${\bf P} = \int_{-\infty}^{0} dt \,\frac{1}{\hbar} 
\left[ {\bf \Pi} (0,t) - {\bf \Pi}(0,-\infty) \right]$ is determined by
\begin{equation}
{\bf {\tilde W} P} = ({\bf p}^{\rm st} \otimes {\bf e}^T - {\bf 1}) ({\bf
  1}-\delta_{\chi',\chi_0}) 
- \partial {\bf \tilde W}{\bf p}^{\rm st} \otimes {\bf e}^T .
  \label{eq:Wtildep}
\end{equation}
Here we use the extra condition 
${\bf e}^T {\bf P} = {\bf 0}$, which follows from the definition of $\bf P$, 
the Dyson equation, and ${\bf e}^T {\bf W} = {\bf 0}$. 
The set of matrix equations Eqs.~(\ref{eq:master}) - (\ref{eq:Wtildep}) 
constitute the starting point for all numerical results presented below.
For a systematic perturbation expansion of current $I$ and noise $S$ in 
$\Gamma$, we expand all quantities ${\bf W}$, 
${\bf W}^I$, ${\bf W}^{II}$, $\partial {\bf W}$, $\partial {\bf W}^I$, 
${\bf p}^{\rm st}$ and ${\bf P}$ order by order. 
We remark that for transport in first order (sequential tunneling), the above
expressions simplify, as all contributions involving the derivatives
$\partial {\bf W}$ and $\partial {\bf W}^I$ disappear as a consequence of the 
fact that ${\bf W}$ starts at order $\Gamma$, ${\bf p}^{\rm st}$ at 
$\Gamma^0$, and ${\bf P}$ at $\Gamma^{-1}$.
These derivatives are associated with non-Markovian behavior of the system,
and have not been taken into account in 
Refs.~\onlinecite{sequential} and \onlinecite{thielmann}.
They are absent for first-order transport but are important for second-
and higher-order corrections.
(For a discussion of non-Markovian effects see also Ref.~\onlinecite{braggio}.)
Higher derivatives will not appear for the shot noise even for higher-order 
corrections.

In the following we will focus on the Anderson impurity model, but
 we emphasize that the expressions presented above also cover
more general situations, such as quantum dots or molecules 
with many levels and more general forms of the 
interactions\cite{hettler_prl,caveat}.
\begin{figure}[h]
\centerline{\includegraphics[width=7.5cm,angle=270]{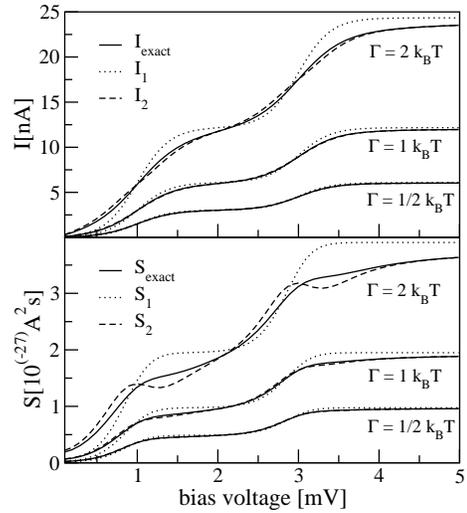}}
\caption{Current $I$ and shot noise $S$ vs bias voltage for $k_{\rm B}T=0.1 {\rm meV}, 
\epsilon_\downarrow=-1.5 {\rm meV}, \epsilon_\uparrow=0.5 {\rm meV}, U=0$, and 
$\Gamma_{\rm L}=\Gamma_{\rm R}=\Gamma/2$. First order (dotted lines) and second
order (dashed lines) are compared to the exact results (solid lines) for
$\Gamma/k_{\rm B}T=0.5, 1, 2$. 
}
\label{fig:is-u0}
\end{figure}

{\bf Results.} --
We recall first the main features of 
first-order (sequential) transport. Both current
and noise increase monotonically with bias voltage displaying plateaus  
separated by thermally broadened steps. The step positions are 
determined by the energy of 
single-charge excitations.
The coupling
parameters set the plateau heights. 

Higher-order processes modify the current and noise in two different ways.
First, they increase the width of the steps, which
effectively is given by the sum of $\Gamma$ and $T$.
Second, they allow for transport in the Coulomb-blockade region at low bias, 
where sequential tunneling is suppressed.
With increasing coupling strength $\Gamma$, second- and eventually also 
higher-order corrections to transport become more and more important.
To illustrate the validity range of our second-order perturbation expansion
we first consider the non-interacting limit, $U=0$,
since in this case exact results\cite{blanter,hershfield} are available 
for the current
$I_{U=0}= e/h \int d\omega  \sum_{\sigma} \tau_{\sigma}(\omega)
(f_{\rm L}(\omega)-f_{\rm R}(\omega) )$ and shot noise 
$S_{U=0}= 2e^2/h \int d\omega \sum_{\sigma} \{ \tau_{\sigma}(\omega)
[f_{\rm L}(\omega)(1-f_{\rm L}(\omega)) + 
f_{\rm R}(\omega)(1-f_{\rm R}(\omega))] 
 + \tau_{\sigma}(\omega)(1- \tau_{\sigma}(\omega))
[f_{\rm L}(\omega)-f_{\rm R}(\omega)]^2 \} $
with $\tau_{\sigma}(\omega)=\Gamma_{\rm L}\Gamma_{\rm R}
 /[(\omega-\epsilon_{\sigma})^2+(\Gamma/2)^2]$.
In Fig.~\ref{fig:is-u0} we compare current and noise in first and second-order
with the exact results.
We choose a symmetric bias voltage ($\mu_{\rm L}=-\mu_{\rm R}=eV/2$).
Outside the Coulomb-blockade regime, the second-order corrections (dashed
lines) to sequential tunneling (dotted lines) start to become important for 
$\Gamma/k_{\rm B}T < 0.5$.
As long as $\Gamma/k_{\rm B}T < 1$, the exact curves (solid lines) are perfectly
reproduced by second-order perturbation theory, while the sequential-tunneling
results clearly deviate.
For $\Gamma/k_{\rm B}T=2$ third-order contributions start to play a role, at least
for the noise, where unphysical non-monotonicities arise around the steps.
We, therefore, restrict ourselves in the following
discussion to $\Gamma \le k_{\rm B}T$. 
 
\begin{figure}[h]
\centerline{\includegraphics[width=6.5cm,angle=270]{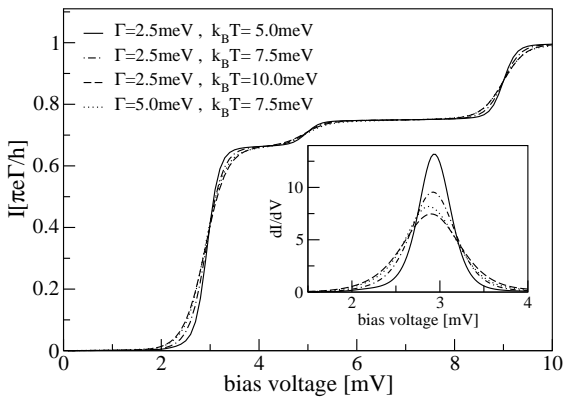}}
\caption{Current $I$ and conductance $dI/dV$ (inset) vs biasvoltage for
 $\epsilon_\downarrow=-1.5 {\rm meV}, \epsilon_\uparrow=0.5 {\rm meV}, U=4 {\rm meV}$ and 
$\Gamma_{\rm L}=\Gamma_{\rm R}=\Gamma/2$ for various values
of $\Gamma$ and $k_{\rm B}T$. The broadening of the
first step due to $\Gamma$ and $k_{\rm B}T$ is shown in the inset. The 
dashed and dotted curves with the same $(\Gamma+k_{\rm B}T)$ have about the 
same width.}
\label{fig:current}
\end{figure}
The elastic co-tunneling processes which do
not change the dot state or its energy
allow for an electron exchange with the reservoirs via an 
intermediate virtual state. This leads to a finite linear conductance 
$G=dI/dV|_{V=0}$. Also the noise persists at zero bias, $S=4k_{\rm B}T G$, consistent with the 
equilibrium fluctuation-dissipation theorem (FDT).
In the Coulomb blockade regime the FDT can be extended to 
non-equilibrium\cite{loss} and  takes the form 
$S^{(2)}(V)/2eI^{(2)}(V)=\coth(eV/2k_{\rm B}T)$. 
Our theory satisfies this relation, however, we stress that it holds 
only in the regime of purely {\it elastic} co-tunneling processes. 
In Fig.~\ref{fig:current}
we show the current $I$ normalized to $\pi e \Gamma/h$ for the same set
of energy parameters as in Fig.~\ref{fig:is-u0} but with a finite  interaction
$U=4 {\rm meV}$. Since the bias is applied symmetrically, the dot preferably
occupies the state with spin $\downarrow$  (Coulomb-blockade) until it can be 
emptied due to first order hopping processes around $3 {\rm mV}$ (first step). 
Further steps arise around $5 {\rm mV}$ and  $9 {\rm mV}$ due to the double occupied dot 
state. This parameter set is similar to the experimental situation of
Ref.~\onlinecite{kouwenhoven} for a quantum dot with occupation
$N=2$\cite{mapping}.
In Fig. 3 of that paper, a conductance step is observed inside 
the Coulomb blockade diamond, 
that is attributed to inelastic co-tunneling processes.  
For our model  one expects 
this inelastic co-tunneling feature in the conductance
at a bias of  $\epsilon_{co}/e=(\epsilon_\uparrow - \epsilon_\downarrow)/e 
= 2 {\rm mV}$.
This feature is hardly noticable
in the conductance plot of the inset in Fig.~\ref{fig:current}, because
the coupling $\Gamma$ is relatively weak and the energy $\epsilon_{co}$
is fairly close to the sequential-tunneling energy.
However, the inelastic co-tunneling processes can clearly be observed 
in the shot noise and the Fano factor $F=S/2eI$ discussed below.

We note that the dashed and dotted curves in Fig.~\ref{fig:current} 
with same total sum  $(\Gamma+ k_{\rm B}T)$ almost lie on top of each other. 
The differential conductance plot (inset) 
shows that the temperature effect is a little stronger:
the dashed curve with the highest temperature has the lowest peak.
The FWHM of the conductance peaks is between  
$0.5{\rm mV} -0.8{\rm mV} \sim 6 (\Gamma+ k_{\rm B}T)$, as compared
to $5.44  k_{\rm B}T$ for pure sequential tunneling\cite{kouwenhoven}.
We also note a shift of the peak position from the sequential-tunneling 
value $3 {\rm mV}$ to somewhat lower bias voltages, 
indicating a renormalization of the level positions.
\begin{figure}[h]
\centerline{\includegraphics[width=7cm,angle=0]{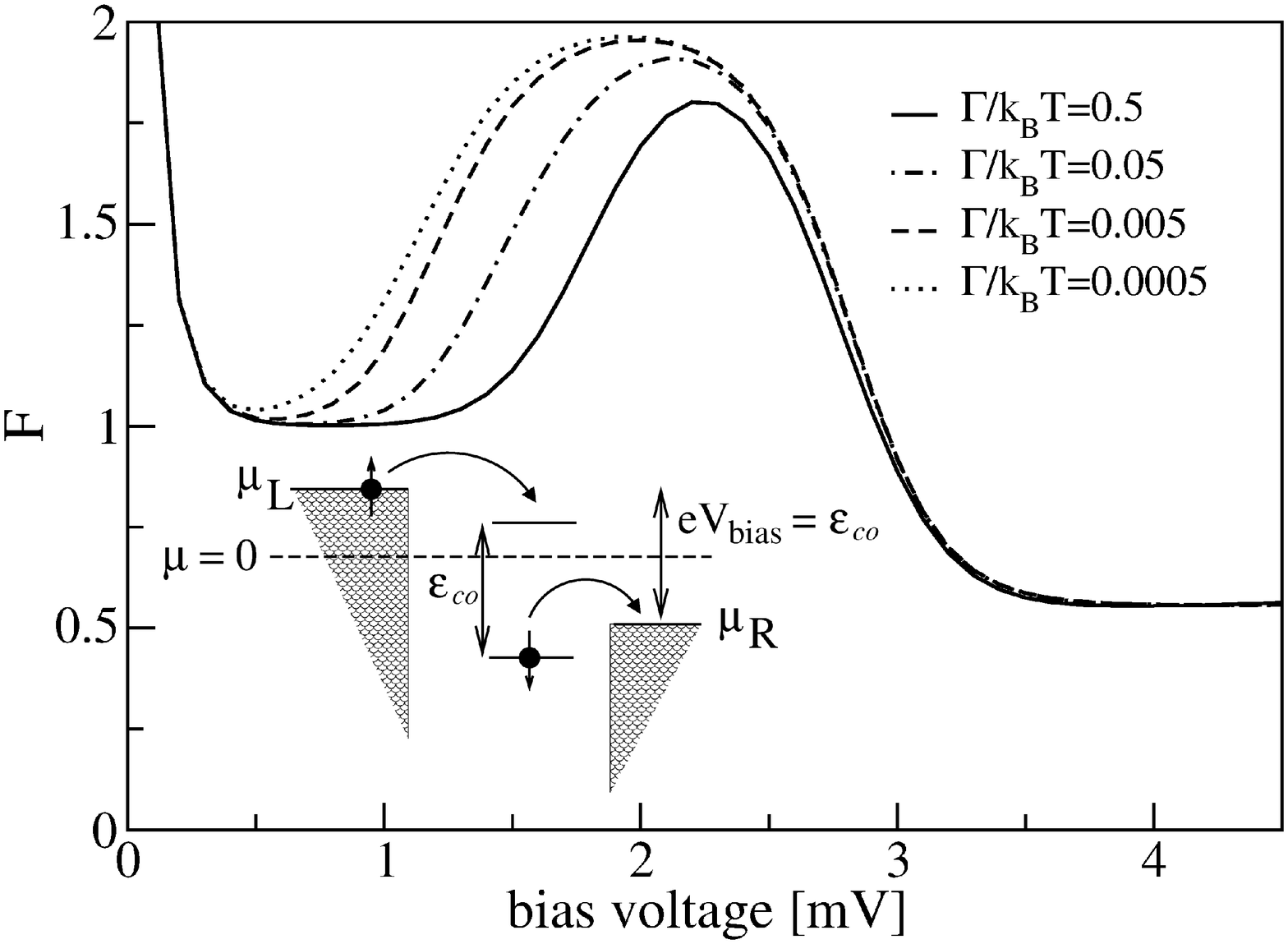}}
\caption{Fano factor $F=S/2eI$ vs bias voltage for the same parameter set
as Fig.~\ref{fig:current}
but fixed temperature $k_{\rm B}T=0.1 {\rm meV}$ and various $\Gamma$.  
Inelastic co-tunneling leads to a super-Poissonian Fano factor 
if the bias is larger than $\epsilon_{co}/e=2 {\rm mV}$. First order processes 
may also lead to a super-Poissonian value at a scale
$2\epsilon_\uparrow/e=1 {\rm mV}$. The crossover between these energy scales 
runs over three orders of magnitude in the coupling.
Outside the Coulomb blockade regime the first order results are 
recovered already at $\Gamma/k_{\rm B}T \sim 0.1$.
The inset shows a sketch of the transport situation at $eV_{\rm bias} = \epsilon_{co}$. }
\label{fig:fano}
\end{figure}

The Fano factor $F=S/2eI$ for a fixed temperature and a sequence of coupling 
constants $\Gamma$, covering three orders of magnitudes, is shown in 
Fig.~\ref{fig:fano}.
At low bias, the Fano factor varies\cite{loss} as $\coth (eV/2k_BT)$ until
it reaches the value 1, as expected for uncorrelated systems.
For bias voltages around the spin-flip excitation energy 
$\epsilon_{co} = 2 {\rm meV}$, the Fano factor becomes super-Poissonian\cite{loss}, 
${F>1}$.
Once sequential tunneling becomes dominant (at a bias $\ge 3 {\rm mV}$), 
it drops to values between 1 and 1/2.

The super-Poissonian Fano factor appears for bias voltages at which the 
spin-$\uparrow$ level acquires some finite occupation probability, either due
to inelastic spin-flip co-tunneling, or due to sequential tunneling processes
that are exponentially suppressed but, for the chosen parameters, still finite.
The enhancement of the noise is mainly described by the second and third term 
of Eq.~(\ref{eq:noise_ptilde}).
It describes bunching of the transfered $\uparrow$-electrons during the time
when this transport channel is not blocked by the dot being occupied with a
$\downarrow$-electron.
Spin flip processes leading to super-Poissonian noise have also been
discussed in Ref.~\onlinecite{ferro_leads} 
in the case of ferromagnetic leads.
Both the position and the height of the peak in the Fano factor depends on all 
system parameters.
In Fig.~\ref{fig:fano} we study the dependence on the ratio $\Gamma/k_{\rm B}T$.

With reduced coupling strength $\Gamma$, 
the peak increases and moves towards lower bias.
For $\Gamma/k_{\rm B}T=0.0005$ our result (dotted line) coincides with that of 
a pure first-order calculation. 
At larger coupling, though, the second-order terms lead to a severe
modification of the result because the rates in the Coulomb-blockade regime
depend only algebraically and not exponentially on the energy.
We emphasize that, since the peak is close to the onset of sequential 
tunneling, an analysis purely based on co-tunneling processes\cite{loss} would
not be sufficient either. 
The range of $\Gamma/k_{\rm B}T$ ratio over which second-order dominated 
co-tunneling crosses over to sequential tunneling is rather large, i.e.,
the Fano factor provides a sensitive measure of the coupling 
strength. For some range of the ratio  
$\Gamma/k_{\rm B}T$ the inelastic co-tunneling effects 
on the shot noise can be 
measured experimentally\cite{noise_level}.
The importance of the second-order processes for the peak of the Fano factor
contrasts with the situation at larger bias beyond
the sequential tunneling threshold when second-order corrections only become
noticable for $\Gamma/k_{\rm B}T \sim 0.1 $.
    
In summary, we presented a theory of current and shot noise 
within a diagrammatic technique that includes higher-order tunneling processes
in the coupling of a quantum dot to metallic electrodes. 
As an example, we studied current and noise for an Anderson impurity model 
with finite spin splitting.
We showed that especially the steps and the Coulomb-blockade regions
are strongly affected by second-order processes and 
provide additional information compared to first order. 
Spin-flip processes of different
origin in first and second order lead to super-Poissonian noise in the 
Coulomb-blockade regime. This provides a spectroscopic tool 
to characterize the dot-electrode couplings by
a combined measurement of current and shot noise.

{\em Acknowledgments.}
We profitted from discussions with J. Aghassi,
A. Braggio, R. Fazio, G. Johansson, and J. Martinek,
and acknowledge financial support by the DFG via the Center for 
Functional Nanostructures, SFB 491 and GRK 726.


\begin{thebibliography}{99}
\bibitem{caveat} In systems with degeneracies not related to symmetries of the
Hamiltonian, a generalization of our approach to off-diagonal density
matrices might be necessary.
\bibitem{mapping}
This is because the spin-split single impurity Anderson model can be mapped
onto a two-level Anderson model\cite{bulka,thielmann_cm,ferro_leads} 
for the temperature and bias range of interest.
\bibitem{noise_level} In Ref. \onlinecite{nauen} shot noise of size $\sim
  10^{-28} A^2 s$ was measured.
\end{thebibliography}
\end{document}